\documentclass[final,times,twocolumn]{elsarticle}
\usepackage{graphicx,amssymb,amsmath,multirow,dcolumn,bm,latexsym,soul}
\usepackage{hyperref,wrapfig}
\usepackage{xcolor}
\usepackage[british]{babel}

\journal{Physics Letters B}
\begin{document}
\begin{frontmatter}

\title{Evolution of current-carrying string networks}
\author[inst1]{J. R. C. C. C. Correia}\ead{Jose.Correia@helsinki.fi}
\author[inst2,inst3]{C. J. A. P. Martins}\ead{Carlos.Martins@astro.up.pt}
\author[inst2,inst4]{F. C. N. Q. Pimenta}\ead{francisco.n.pimenta@gmail.com}
\address[inst1]{Department of Physics and Helsinki Institute of Physics, PL64, FI-00014 University of Helsinki, Finland}
\address[inst2]{Centro de Astrof\'{\i}sica da Universidade do Porto, Rua das Estrelas, 4150-762 Porto, Portugal}
\address[inst3]{Instituto de Astrof\'{\i}sica e Ci\^encias do Espa\c co, CAUP, Rua das Estrelas, 4150-762 Porto, Portugal}
\address[inst4]{Faculdade de Ci\^encias, Universidade do Porto, Rua do Campo Alegre, 4150-007 Porto, Portugal}
\begin{abstract}
Cosmic string networks are expected to form in early Universe phase transitions via the Kibble mechanism and are unavoidable in several Beyond the Standard Model theories. While most predictions of observational signals of string networks assume featureless Abelian-Higgs or Nambu-Goto string networks, in many such extensions the networks can carry additional degrees of freedom, including charges and currents; these are often generically known as superconducting strings. All such degrees of freedom can impact the evolution of the networks and therefore their observational signatures. We report on the results of $2048^3$ field theory simulations of the evolution of a current-carrying network of strings, highlighting the different scaling behaviours of the network in the radiation and matter eras. We also report the first numerical measurements of the coherence length scales for the charge and current and of the condensate equation of state, and show that the latter mainly depends on the expansion rate, with chirality occurring for the matter era. Qualitatively, the fact that the matter era is the optimal expansion rate for these networks to reach scaling is in agreement with recent analytic modeling.
\end{abstract}

\end{frontmatter}

\section{Introduction}\label{intr}

Topological defects are expected to form in early Universe phase transitions, due to symmetry breaking processes \cite{Kibble:1976sj}. Such phase transitions are ubiquitous in Beyond the Standard Model particle physics theories \cite{Jeannerot1,Jeannerot2}, and often lead to one-dimensional line-like cosmic strings. These typically do not overclose the Universe but can produce astrophysical signatures detectable or constrained by current \cite{Planck:2013mgr,LIGO,NANOGrav:2023hvm,EPTA} or future facilities \cite{LISA,Weltman:2018zrl}. 

Credible and accurate cosmic strings observational constraints require a description of the evolution of a string network throughout cosmic history. This involves both analytic modeling \cite{Kibble85,Bennett86,Vachaspati87,Kibble91,Austin93}---leading to the velocity-dependent one-scale (VOS) model \cite{Martins:1996jp,Martins:2000cs,Book}---and Nambu-Goto \cite{Albrecht,Allen,Bennett,Ringeval,Blanco-Pillado:2011egf} and field theory \cite{Vincent,Moore,Bevis,Daverio,Hindmarsh,Correia:2019bdl} numerical simulations. A shortcoming of these approaches so far is that they rely on the simplest (Abelian-Higgs or Nambu-Goto) cosmic strings, while physically realistic networks should carry charges and/or currents \cite{Witten}. 

As no field exists in isolation, any field that underpins string configurations will generically couple to other existing bosonic or fermionic fields. Such couplings lead to strings with extra degrees of freedom, e.g. a conserved current and a modified internal structure, dynamics and therefore, observational footprints \cite{Rybak:2017yfu}. Bosonic condensates are expected from Higgs-like degrees of freedom \cite{Peter3} and fermionic fields can provide propagating zero modes along the worldsheet \cite{Peter1}. When the symmetry underpinning the current sector corresponds to a gauged $U(1)$ these objects become superconducting, as the broken electromagnetic gauge invariance is a tell-tale sign of superconductivity. These networks may also be a source of dark matter \cite{Davis:1988ij,Peter:1993pz,Auclair:2020wse} and other novel astrophysical signatures \cite{Vilenkin:1986zz,Vachaspati:2009kq,Theriault:2021mrq}.

The charge-velocity-dependent one-scale (CVOS) model, an extension of the original VOS model accounting for generic charges and currents, has been recently developed \cite{Currents1,Currents2}. This aims to preserve, as far as possible, the conceptual simplicity of the VOS model, while phenomenologically incorporating the additional physical processes which generic current-carrying strings are expected to include \cite{Peter92a,Peter92b,Carter95,Carter01,Copeland05}. Here we report on field theory numerical simulations of such networks, and present a first qualitative comparison with the analytic model.

Despite the recent development of the CVOS model, any such model is incomplete until tested and calibrated by high-resolution numerical simulations. There are very few attempts to simulate strings or networks with currents \cite{Lemperiere:2003yt,Battye:2021kbd,Battye:2022mxi}, let alone characterize their scaling properties. Here we present the first results of full numerical simulations allowing for charges and current on the strings. Broadly speaking the simulations agree with the expectations inferred from the model, although a more detailed and robust comparison is left for future work. We rely on our multi-GPU accelerated code \cite{Correia:2020yqg}, also used to study multiple interacting string networks \cite{Correia:2022spe}, and appropriately modify it to study current-carrying string networks. 

\section{Bosonic current-carrying strings}\label{prac}

The few previous attempts to simulate current-carrying strings \cite{Lemperiere:2003yt,Battye:2021kbd,Battye:2022mxi} are all based upon field theory simulations. Here we adopt the model of \cite{Saffin:2005cs}, with a Lagrangian density and potential
\begin{equation}
\mathcal{L} = |D_{\mu}\phi|^2 -\frac{1}{4} F^{\mu\nu}F_{\mu\nu}+|\mathcal{D}_{\mu}\sigma|^2 -\frac{1}{4} G^{\mu\nu}G_{\mu\nu} -V(\phi,\sigma) \, , 
\end{equation}
\begin{align}
    V(\phi,\sigma) &= \frac{\lambda_\phi}{4}(|\phi|^2 -\phi_0^2)^2 +\frac{\lambda_\sigma}{4}(|\sigma|^2 - \sigma_0^2)^2 \nonumber \\
                     &-\kappa (|\phi|^2 - \phi_0^2)(|\sigma|^2 - \sigma_0^2) \,,
\end{align}
comprising two $U(1)$ symmetries, each with a complex scalar field and a corresponding gauge field and gauge field strengths, $\phi,A_\mu,F_{\mu,\nu}$ and $\sigma,B_\mu,G_{\mu,\nu}$, respectively. The gauge field strengths and covariant derivatives have the usual definitions
\begin{equation}
    F_{\mu\nu} = \partial_{\mu}A_{\nu} - \partial_{\nu}A_{\mu}\,, \qquad G_{\mu\nu} = \partial_{\mu}B_{\nu} - \partial_{\nu}B_{\mu}\, ,
\end{equation}
\begin{equation}
    D_{\mu} = \partial_{\mu} -ie_\phi A_{\mu}\,, \qquad \mathcal{D}_{\mu} = \partial_{\mu} -ie_\sigma B_{\mu}\, ,
\end{equation}
where $\lambda_{\phi,\sigma}$ are scalar couplings, $e_{\phi,\sigma}$ are gauge couplings and $\kappa$ is the coupling between the two scalar fields. In our previous work \cite{Correia:2022spe} we have chosen parameters such that $0 < \kappa < \frac{1}{2}\sqrt{\lambda_\phi \lambda_\sigma}$, which led to $U(1)\times U(1)$ multitension networks. Here we instead explore the regime where the $\sigma$ $U(1)$ symmetry remains unbroken far away from the strings, while the $\phi$ $U(1)$ sector is broken and leads to the formation of current-carrying cosmic strings. In other words, in this work we still simulate the same model, but focus on a different parameter space region thereof.

Without loss of generality we work with $e_\sigma=0$ to further localize the condensate to the string worldsheet, and simplify initial condition generation. Strictly speaking, these are no longer superconducting strings, but are still current-carrying strings. This allows us to identify $\sigma$ as the current carrier field, and $\phi$ and $A_\mu$ as our Higgs fields, responsible for the string itself. As described in \cite{Saffin:2005cs}, the existence of current-carrying strings requires $\kappa < -\frac{1}{2} \sqrt{\lambda_\phi \lambda_\sigma}$, which allows minima of the potential to exist at $(|\phi_0|^2 -\frac{2\kappa}{\lambda_\phi} |\sigma_0|^2,0)$ and $(0,|\sigma_0|^2 -\frac{2\kappa}{\lambda_\sigma} |\phi_0|^2 )$. 

To ensure stability of the vacuum, we must also have $\lambda_\phi \phi_0^4 > \lambda_\sigma \sigma_0^4$ and it is desirable to have the mass of the condensate outside the string core ($m_{\sigma,+}^2 = -\frac{1}{2} \lambda_|\sigma_0|^2$) larger than its mass inside the string ($m_{\sigma,-}^2 = k|\sigma_0|^2$). In other words, we must ensure $-\frac{1}{2} \lambda_\sigma |\sigma_0|^2 > \kappa |\phi_0|^2$ or $\frac{|\sigma_0|^2}{|\phi_0|^2} < \frac{-2 \kappa}{\lambda_\sigma}$. 

The equations of motion can be obtained by standard variational techniques. We assume a Friedmann-Lemaitre-Robertson-Walker Universe, where the scale factor is given by $a(t)\propto t^\lambda \propto \eta ^ {\lambda/(1-\lambda)}$, and $t$ and $\eta$ are physical and conformal time respectively. The metric signature used throughout this paper will be $(+,-,-,-)$. The equations of motion for the scalar fields are then
\begin{equation}
\ddot{\phi} + 2\frac{\dot{a}}{a}\dot{\phi} = D^jD_j\phi -a^2 \phi \bigg[ \frac{\lambda_\phi}{2} (|\phi|^2 - \phi_0^2) - \kappa (|\sigma|^2 - \sigma_0^2) \bigg] \, , 
\end{equation}
\begin{equation}
\ddot{\sigma} + 2\frac{\dot{a}}{a}\dot{\sigma} = \partial^j \partial_j\psi -a^2 \sigma \bigg[ \frac{\lambda_\sigma}{2} (|\sigma|^2 - \sigma_0^2) - \kappa (|\phi|^2 - \phi_0^2) \bigg] \, ,
\end{equation}
while the gauge fields' is
\begin{equation}
\dot{F}_{0j} = \partial_j F_{ij} -2a^2 e_\phi^2 Im[\phi^* D_j \phi]\, .
\end{equation}
along with Gauss' law for the first sector
\begin{equation}
\partial_i F_{0i} - 2 a^2 e_\phi^2 \Im[\phi^* \dot{\phi}]\phi = 0 \, ,
\end{equation}
and a charge conservation equation for the second sector
\begin{equation}
  \partial_\mu J = \sigma \partial_\mu \sigma = 0 \, .
\end{equation}
Numerically we start the Higgs sector with the same parameters used for our earlier Abelian-Higgs strings simulations ($\lambda_\phi = 2$, $e_\phi = 1$, $\phi_0 = 1$) and with $\lambda_\sigma=10$, $\sigma_0=0.61$ for the second sector, with coupling constant, $\kappa = -3.0$. These guarantee the stability of the vaccum, and also that the condensate mass is larger outside the core.

\section{Numerical implementation}\label{overview}

We follow the standard strategy for discretizing fields on a lattice, applying the principles of lattice field theory \cite{PhysRevD.10.2445} in an expanding background: allowing scalar fields to live in points and gauge fields in links, playing the role of parallel transporters, thus allowing for local $U(1)$ gauge invariance on the lattice \cite{Hindmarsh,Correia:2020yqg,Correia:2022spe}.

Expansion does introduce a resolution bottleneck: as all fields live on a comoving lattice, the constant physical scales of the system (which control the string and condensate widths) will appear to shrink as $a^{-2}$. We therefore use the constant comoving width prescription of \cite{PRS} which modifies the scalar and gauge couplings such that their corresponding widths are constant in comoving coordinates:
\begin{equation}
\lambda_{\phi,\sigma} = \lambda_{\phi,\sigma0} a^{-2(1-\beta)}\,, \quad e_\phi = e_{\phi0} a^{-(1-\beta)}\,, \quad  \kappa = \kappa_0 a^{-2(1-\beta)} \, .
\end{equation}
where the couplings subscripted with $0$ correspond to the physical versions defined above. Following \cite{Lizarraga:2016hpd,Correia:2022spe} $\kappa$ is made to vary exactly as $\lambda_{p,q}$. Setting $\beta=0$ forces constant comoving width, while $\beta=1$ recovers the original equations of motion. The fully discretized equations of motion are the same ones presented in \cite{Correia:2022spe}. 

While there is no unique way to set initial conditions, $\phi$ should have values close to its vacuum expectation value and that $\sigma$ should consist of thermal fluctuations. A simple way to achieve this is to set
\begin{equation}
  \phi = |\phi_0|^2 \exp^{-i \theta_1}\,,\qquad \sigma = |\Upsilon| \exp^{-i \theta_2}
\end{equation}
where $\theta_{1,2} \in [0,2\pi]$ and are drawn from a uniform distribution, while $\Upsilon$ is drawn from a normal distribution with a standard deviation of $0.01$ and a null mean. This mimics having the second scalar field close to the minimum of the potential, with suitably small thermal fluctuations. All gauge fields are set to zero as are the conjugate momenta. 

Merely setting these initial conditions at conformal time $\eta_{0}=1.0$ does not suffice. In some cases (roughly half the times in the radiation era), a "blob-shaped" instability will form, where the string radius grows at a point, and up to horizon size. Assuming that we are seeing the pinching instabilities reported in \cite{Battye:2021kbd,Battye:2022mxi}, it should be related to the balance between string curvature and current on the strings.

Specifically, the aforementioned references reported that vorton loops whose radii were sufficiently large for a given current (and therefore a given current coherence length), would not develop this intrinsic instability. Our goal here is to develop initial conditions that evade the formation of such instabilities, by establishing a hierarchy between string curvature and fluctuations of $\sigma$, which seed current. We have tested that these instabilities are related to the curvature radius by increasing the expansion rate to reduce the number of simulations where "blobs" form. With an expansion rate of $\lambda=0.75$ no instability appears. We thus modify our initial conditions with the hierarchy of fluctuations $l_\sigma < l_\phi \leq d_H$, where $d_H$ is the size of the horizon and $l_{\sigma,\phi}$ are the typical size of fluctuations in $\phi$ and $\sigma$, respectively. Without further modifying the initial conditions of $\sigma$, we start the simulation at $\eta_{0}=5.0$ and apply nearest neighbors smoothing ($40$ times) and damping (setting $\lambda=0.9$) to $\phi$ until time $\eta=6.0$ and subsequently diffusive cooling from $\eta \in [6,10]$, with timesteps of size $\delta t=1/30$. This ensures that at $\eta=10.0$, we start with a network of strings with mean string separation $\xi \approx 10$ and fluctuations of $\sigma$ of order lattice spacing, satisfying the above condition. The diffusive cooling equations to be applied are
\begin{equation}
  \dot{\phi} = D^jD_j\phi -\frac{\lambda_{\phi0}}{2} (|\phi|^2 - \phi_0)\phi - \kappa (|\sigma|^2 - \sigma^2_0) \phi
  \end{equation}
  \begin{equation}
  \dot{F}_{0j} = \partial_j F_{ij} -2 e_{\phi0}^2 {\bf \Im}[\phi^* D_j \phi]\,
\end{equation}
Note how the potential assumed here takes into account the existence of $\sigma$, thus allowing for strings to form with the correct string radius, but $\sigma$ itself is not updated. This is so neither damping nor diffusion drive $\sigma$ too close to zero, to disallow the formation of a condensate a posteriori.

\begin{figure*}
\begin{center}
  \includegraphics[width=0.49\textwidth]{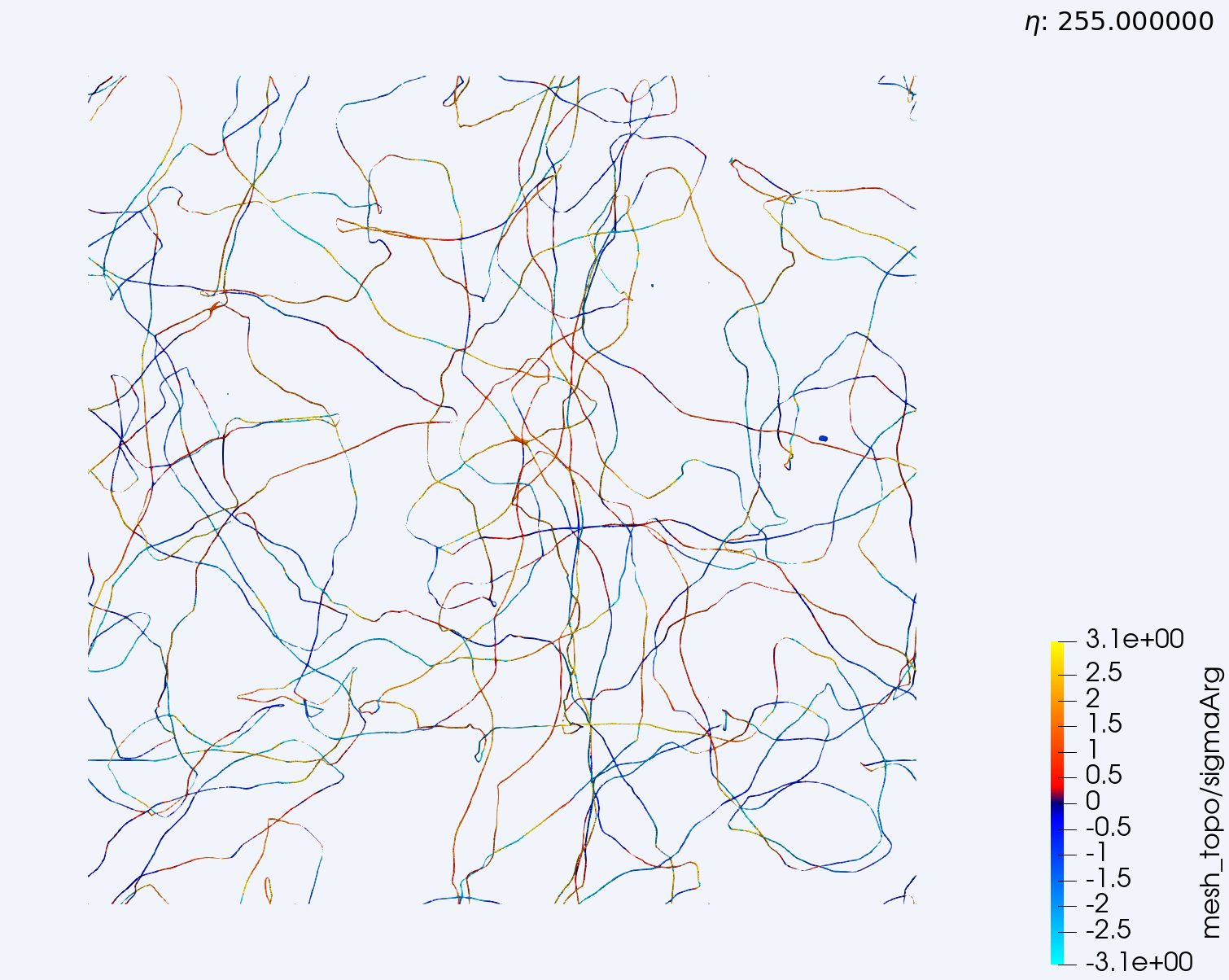}
  \includegraphics[width=0.49\textwidth]{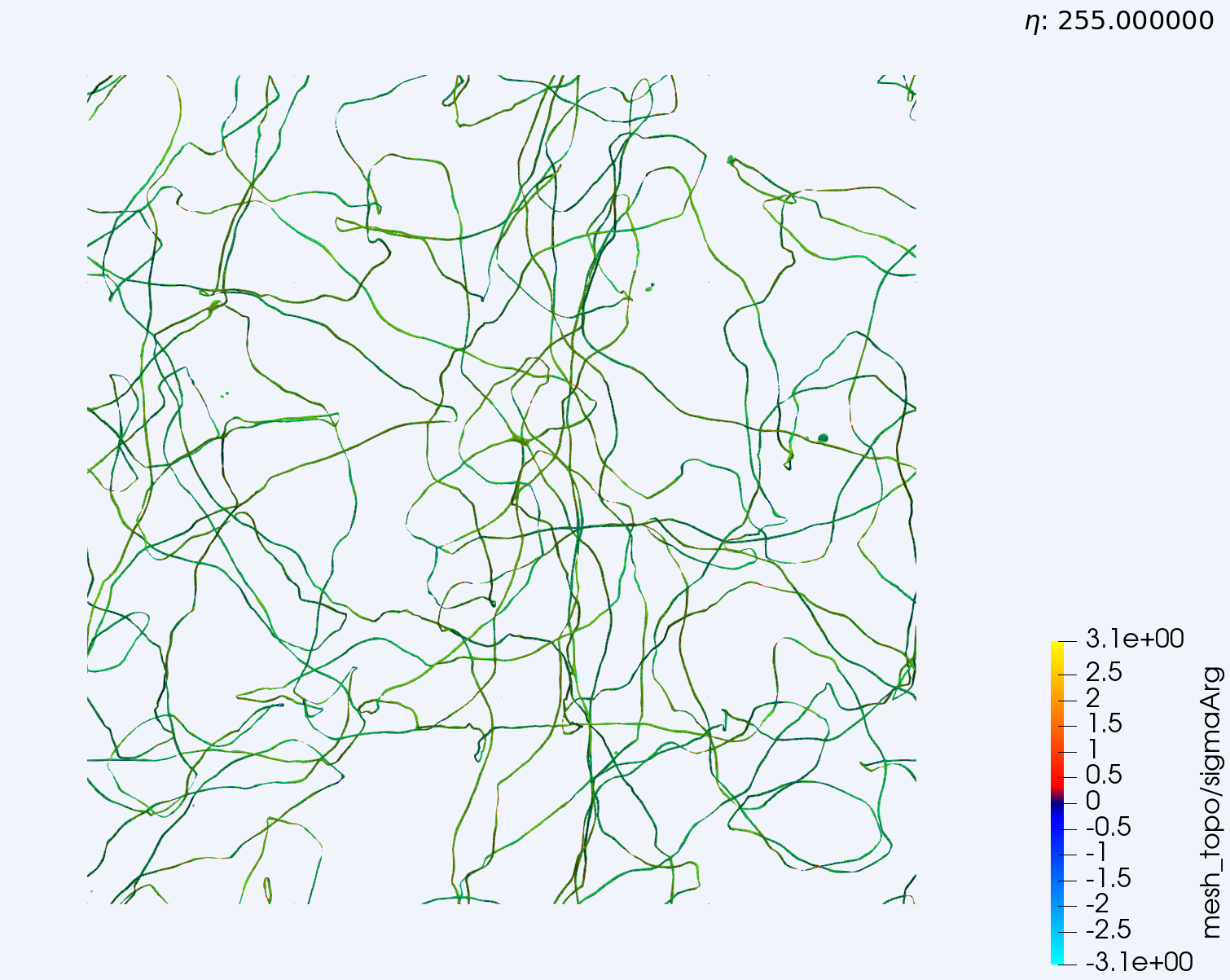}
\end{center}
\caption{Visualization isosurfaces showing the presence of strings and the presence of a condensate wrapped around them. On the left we present iso-surfaces of $|\phi|=0.7$ colored by the phase of the condensate degree of freedom, $\sigma$. On the right we show isosurfaces of $|\sigma|=0.4$ colored in green. This is a radiation-epoch simulation, at conformal time $\eta=255$, roughly half-a-light time crossing for a size $1024^3$, $\Delta x = 0.5$ lattice. Both animations can be found online \cite{zenodo}.\label{figure1}}
\end{figure*}

Cosmological evolution then proceeds until half a light crossing time, $\eta_{final}$. We use $2048^3$ lattices with lattice spacing $\delta x = 0.5$, so our final conformal time is $\eta_{final}=512$. In the first few timesteps, fluctuations in $\sigma$ are unstable \cite{Witten} and will eventually source a non-zero condensate around the string core. The presence of this condensate can be seen on the isosurfaces of $|\sigma| = 0.4$ in the right-hand-side panel of Fig. \ref{figure1}, in a radiation epoch run at half-a-light-crossing $\eta=255$ (lattice size $1024^3$, $\Delta x = 0.5$). Another telltale sign of its presence is the change of phase of $\sigma$ along the string, which occurs over some current coherence length scale to be further discussed below. The presence of the current can then be seen by coloring isosurfaces of $|\phi|=0.7$ with the argument of $\sigma$ in an interval of $[-\pi,\pi]$. This can again be seen on figure \ref{figure1}, left-hand-side panel. We note as well that a Y-junction can be seen in this visualization, in accordance with \cite{Fujikura:2023lil}, albeit it does not seem to be dynamically relevant or statistically important.

In order to fully characterize the late-time state of the network, we need estimators of string length, from which we can extract the mean string separation, and of the root mean square velocity. As in previous work, the mean string separation $\xi$ is defined via the total length of string $L$ in the simulation box,
\begin{equation}
\xi = \sqrt{\frac{ \mathcal{V} }{L}}= \sqrt{\frac{ \mathcal{V} }{\sum_{ij,x} W_{ij,x}}}\,, 
\end{equation}
where $\mathcal{V}$ is the comoving lattice volume and $W_{ij,x}$ is the winding of the field $\phi$ at site $x$ on a plaquette oriented over directions $i$ and $j$. This winding is computed with the prescription of \cite{Kajantie:1998bg}, divided by $2\pi$ to yield an integer, and then multiplied by the lattice spacing $\delta x$. The summation of the total length made up of $\delta x$ segments is then corrected by $\pi/6$ to compensate the taxicab geometry of strings. We remark that we merely add all of these contributions without explicitly separating bound states, although these were found to be statistically insignificant when compared to winding one strings in radiation epoch.

For the root mean square velocity of the field $\phi$ responsible for the strings, we rely on its local gradients and conjugate momenta
\begin{equation}
\label{eq:defvPhiSS}
\overline{v}^2_{\phi} = \frac{2R}{1+R}\,,
\end{equation}
\begin{equation}
\label{eq:defRSS}
R = \frac{\sum_x  |\Pi|^2  f(\phi, \sigma)}{\sum_{x,i} |D^+_{x,i} \phi|^2  f(\phi, \sigma) }\,,
\end{equation}
with $f(\phi,\sigma)$ as a weight function, meant to localize the estimator around strings. For the purposes of the present work, two choices are considered. The first uses the absolute value of the current-carrier field
\begin{equation}
f(\phi, \sigma) = |\sigma|^2\,,
\end{equation}
while the second smoothly interpolates from unity to zero, at either string core or far away depending on the value of $|\phi|$; we adopt the sigmoid-like function,
\begin{equation}
f(\phi, \sigma) = \frac{1}{1+\exp^{10(|\phi| - \delta_\phi)}}\,,
\end{equation}
where the mid-point is chosen to be at $\delta_\phi = m_\phi^{-1}$. The factor of 10 can be changed at will to control how sharp this function transitions between its limits. Based on our choices on isosurfaces of $\phi$, our choice makes the weight yield a value of 0.88 at $|\phi|^2 = 0.6$. A more detailed discussion of this choice is left for future work.

We can also define mean square charge and current
 \begin{equation}
  \overline{Q}^2,\overline{J}^2 = \frac{ \sum_x \Im(\sigma^* \partial_{0,i} \sigma)^2 f(\phi, \sigma) }{ \sum_x \rho f(\phi, \sigma)}
 \end{equation} 
where either charge or current squared are compared with the total energy density. Both can be weighted by the previously defined weight functions $f(\phi, \sigma)$ in order to localize this estimator to the strings themselves. To further simplify notation, we will also drop all overlines from these variables, e.g. $\overline{Q}^2 \rightarrow Q^2$. Following \cite{Currents1,Currents2}, we also consider the auxiliary variables $K=Q^2-J^2$ and $Y=(Q^2 + J^2)/2$. These show whether the current is, on average, propagating along strings as a time, space or light-like degree of freedom, and how close we are to a featureless string network. Abelian-Higgs networks have $K=Y=0$. On the other hand, if $K=0$ and $Y\neq 0$ we have chiral strings (with a light-like current), while having both $K,Y\neq0$ indicates a time-like ($Q^2>J^2$) or space-like current ($J^2>Q^2$).

We can also explicitly compute a local equation of state of the condensate degree of freedom,
\begin{equation}
\omega_\sigma = \frac{Q^2-J^2/3}{Q^2+J^2} \; . 
\end{equation}
This provides a second test of chirality: in that case the charge/current are expected to propagate along the string as massless degrees of freedom, hence $\omega_\sigma=1/3$. We remark that these do not map to the often assumed one-dimensional current $J^2$ and microscopic equation of state $\kappa$ often used on theoretical (Nambu-like) based modelling, such as the CVOS model \cite{Currents1,Currents2}, although they still largely allow us to conclude if our networks are chiral, electric or magnetic.

Finally, we can numerically determine the average coherence length scale associated with the charge or current,
\begin{equation}
  l_{Q,J} = \frac{ \sum_x \Im(\sigma^* \partial_{0,i} \sigma)^2 f(\phi, \sigma)}{ \sum_x |\sigma|^4f(\phi, \sigma)}\,.
 \end{equation} 
From here we can posit this coherence length to lead to some mean charge/current separation, which at sufficiently late times, and a scaling network, should grow linearly with conformal time,
\begin{equation}
\xi_{Q,J} \propto \sqrt{ \frac{1}{l_{Q,J}} } .
\end{equation}

In this manuscript we will test whether this coherence length $l_{Q,J}$, or rather its inverse squared, linearly scales as one would expect from scaling quantities.
\begin{figure*}[p]
\begin{center}
  \includegraphics[width=1.0\columnwidth]{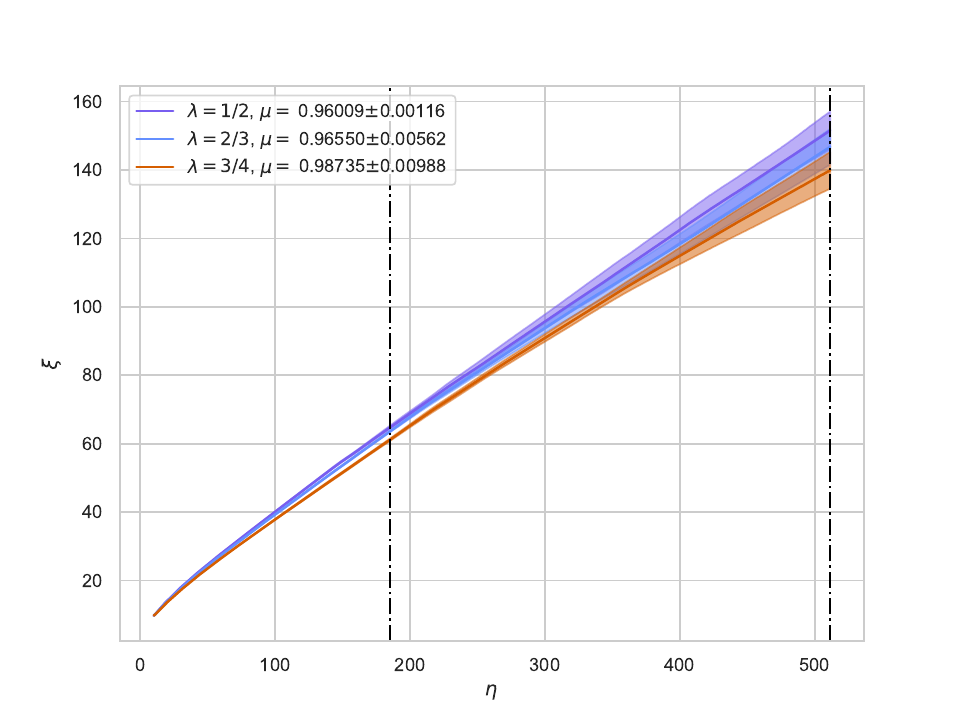}
  \includegraphics[width=1.0\columnwidth]{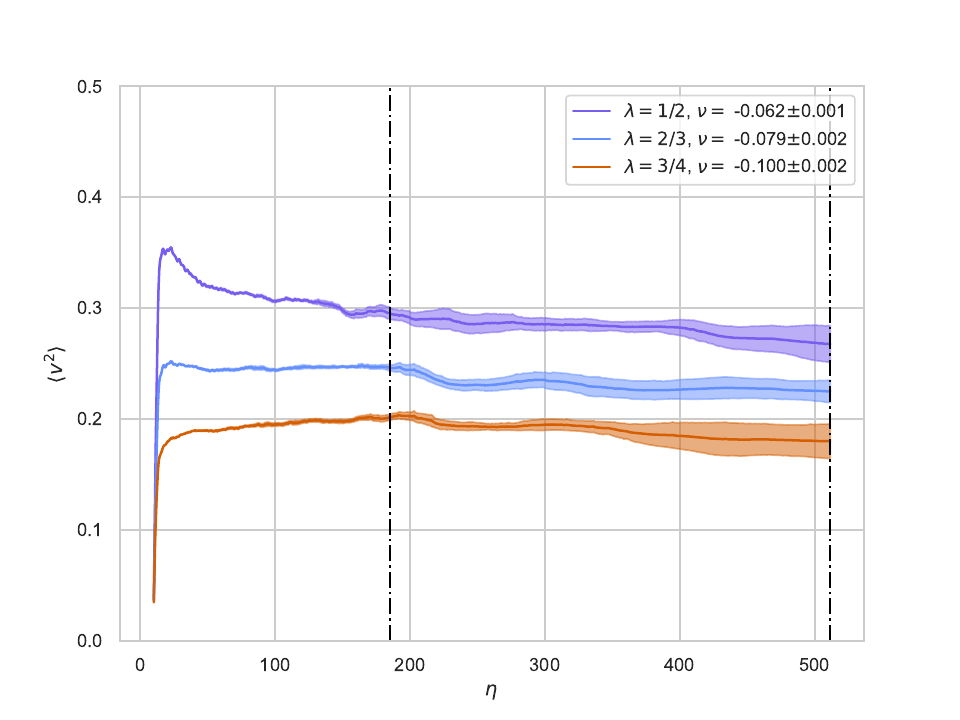}
  \includegraphics[width=1.0\columnwidth]{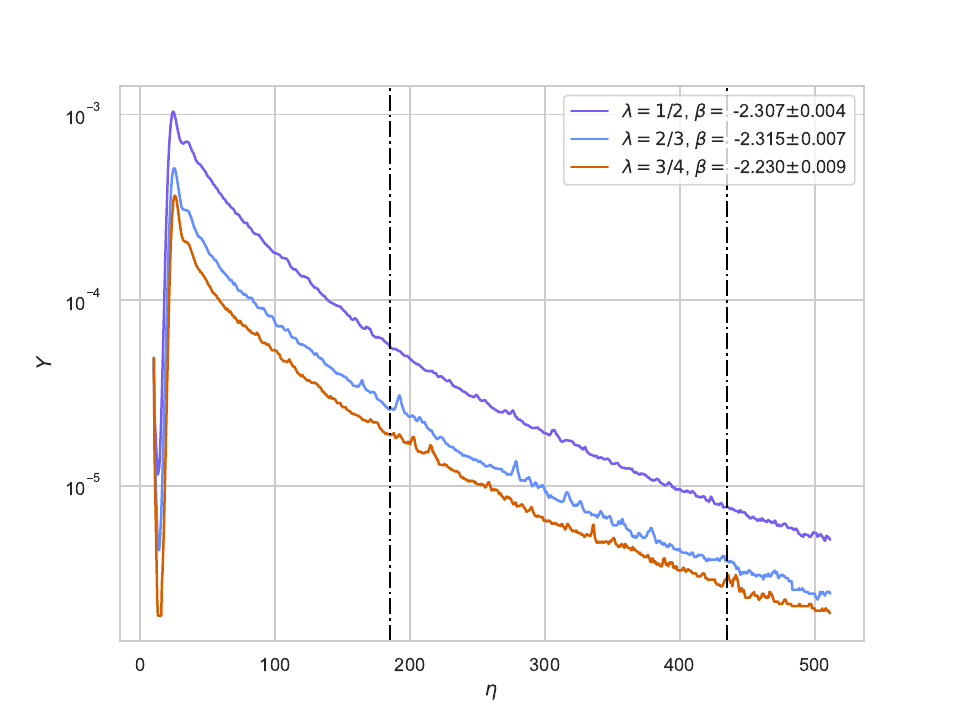}
  \includegraphics[width=1.0\columnwidth]{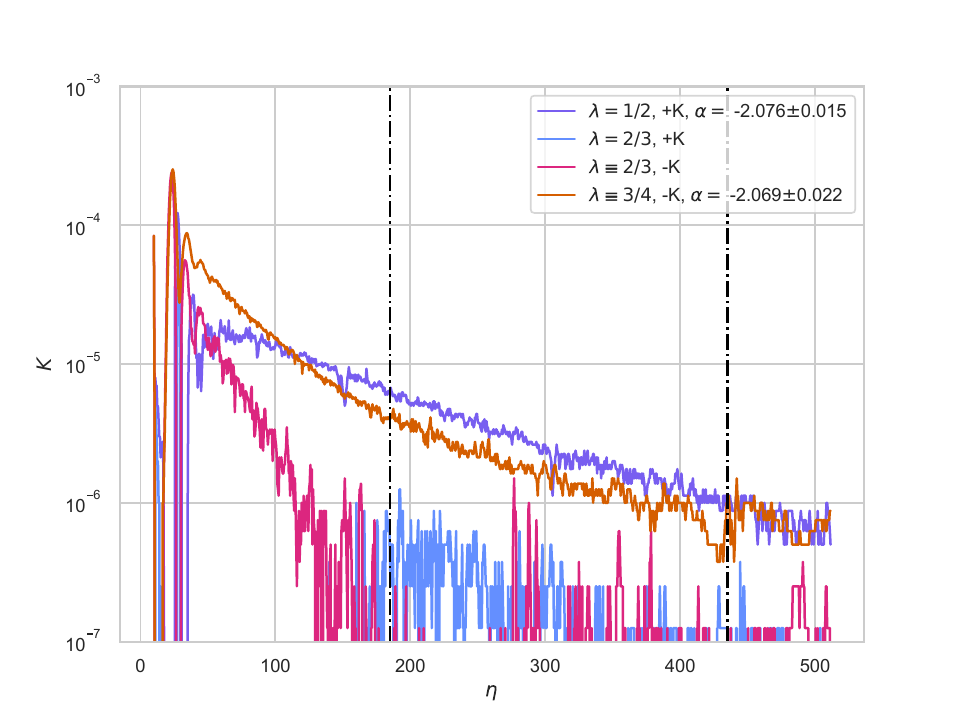}
  \includegraphics[width=1.0\columnwidth]{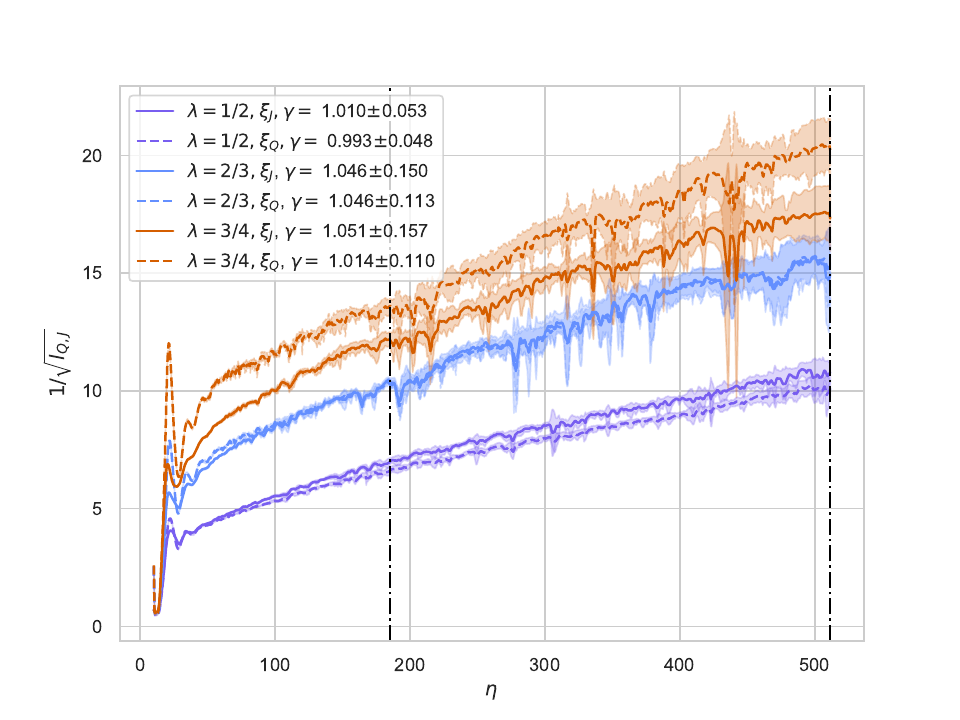}
  \includegraphics[width=1.0\columnwidth]{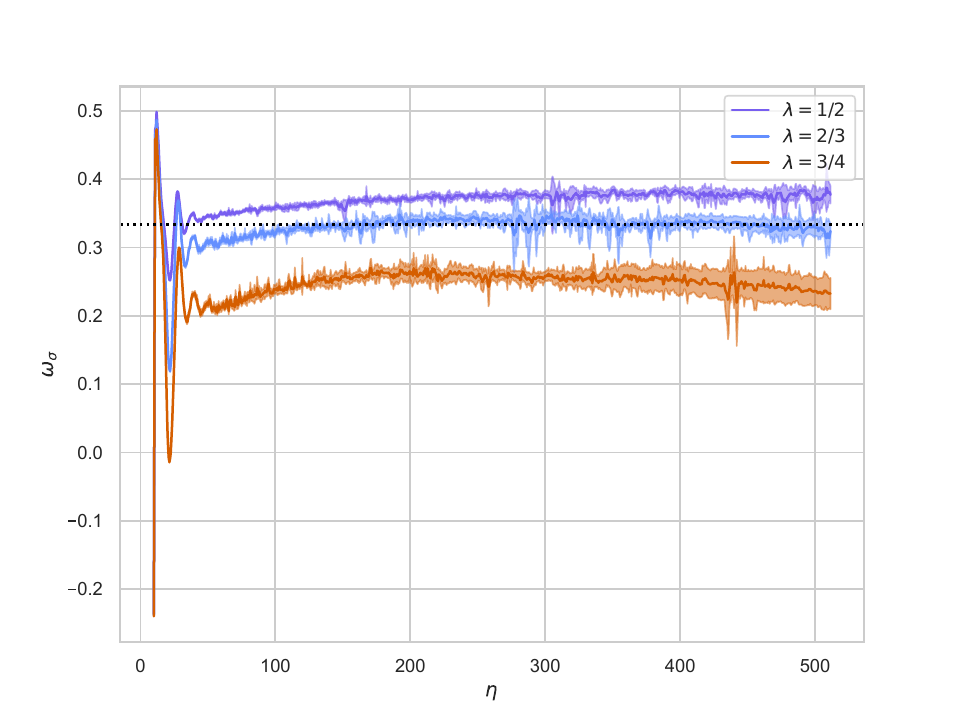}
\end{center}
\caption{\textbf{Top:} mean string separation and root mean square velocity of networks in radiation era (purple) and the matter era (blue) and fast expansion (orange). \textbf{Middle:} average current-charge $Y$ (left) and chirality $K$ (right). Note that for matter epoch we additionally display $-K$ in magenta. \textbf{Bottom:} the mean coherence length of charge and current (left, in dashed and solid lines respectively) and the condensate equation of state (right). Vertical black dash-dotted lines indicate the fitting ranges.  \label{figure2}}
\end{figure*}

\section{\label{Results}Scaling properties of current-carrying strings}

We begin with the canonical network diagnostics of scaling, the mean string separation $\xi$ and velocity $\langle v^2 \rangle$, and fit the simulations for scaling laws
\begin{equation}
\xi \propto (\eta - \eta_0)^\mu\,,\qquad \langle v^2 \rangle \propto \eta^\nu\,,
\end{equation}
where $\mu$ and $\nu$ should be unity and null, respectively, in standard linear scaling.

In all the fits reported in this section, we choose a fitting range in the latter part of the simulations, such that initial conditions have been erased, but there are still enough strings (or charge or current in them), to provide meaningful information. One sigma statistical uncertainties, from the average of ten $2048^3$ simulations with different initial random seeds, are also reported. We note that the results of the fits will therefore depend on the chosen fitting range.

As can be seen in  the upper panels of  Fig. \ref{figure2} (see also Table \ref{tab1}), the matter era simulations are more consistent with linear scaling than those in the slower (radiation) and faster expansion rates. Admittedly this is not fully clear-cut, since the velocities have a slight decreasing tendency. Although we cannot currently exclude difficulties with velocity estimation as a possible cause, the most likely explanation is that matter era simulations approach the standard scaling faster, while in other cases the presence of the additional degrees of freedom slows it down. It is also apparent that the correlation length approaches scaling faster that the velocity or the charge and current. This is physically plausible, implying that the network first reaches large-scale scaling (i.e., its total energy density becomes a constant fraction of the background density), while on smaller scales one still has energy transfers between the bare string and the charge/current (which would impact the velocity). Simulations with larger dynamic range, enabling a more robust statistical analysis, will be needed to confirm this.

\begin{table}
  \begin{center}
\caption{Scaling exponents of the mean string separation $\xi$ and velocity $\langle v^2 \rangle$, along with the asymptotic averages of $\dot{\xi}$ and $\langle v^2 \rangle$. Fitting range for these quantities is $\eta \in [185,512]$. One sigma statistical uncertainties, from the average of ten $2048^3$ simulations, are also reported.}
\label{tab1}
  \begin{tabular}{ |c|c|c| }  \hline
Epoch & $\mu$ & $\nu$ \\   \hline
    $\lambda=1/2$  & $0.9601\pm0.0012$ & $-0.064\pm0.001$ \\ 
    $\lambda=2/3$  & $0.9655\pm0.0006$ & $-0.079\pm0.002$ \\   
    $\lambda=3/4$  & $0.9874\pm0.0010$ & $-0.100\pm0.002$ \\   \hline
Epoch & $\dot{\xi}$ & $\langle v^2 \rangle$ \\   \hline
   $\lambda=1/2$ & $0.266\pm0.015$ & $0.282\pm0.003$  \\
   $\lambda=2/3$ & $0.252\pm0.016$ & $0.231\pm0.002$  \\  
   $\lambda=3/4$ & $0.240\pm0.016$ & $0.189\pm0.005$  \\  \hline
  \end{tabular}
\end{center}
\end{table}

Supporting evidence comes from measuring the charge and current, cf. the middle panels of Fig. \ref{figure2}. We find that matter era networks quickly reach the  chiral limit, wherein $K$ is consistent (within our numerical precision) with zero. In the radiation era the string network is mildly electric with non-vanishing $Q$ and $J$ of order $10^{-4}$,  $Y,K \neq 0$ and $Q>J$, the opposite occurring for the fast expansion rate (the network is mildly magnetic). Note that matter epoch networks are not quite in the Nambu-Goto (or Abelian-Higgs) limit, since $Y \neq 0$. 

We can also compute the scaling exponents for $K$ and $Y$, assuming
\begin{equation}
|K| \propto \eta^\alpha\,,\qquad Y \propto \eta^\beta \,,
\end{equation}
which are reported in Table \ref{table2} (with the obvious exception of $\alpha$ in the matter era). For the radiation and fast expansion epochs, both $|K|$ and $Y$ exhibit power-law decays with negatively exponents, which moreover are very similar in the case of $\alpha$, and slightly less so for $\beta$ (in all three epochs). Note that in the considered fitting range $K$ is consistent with null in the matter epoch (up to numerical noise), which signals chirality. In the fast expansion era, the negative value of $K$ implies that these strings are fully magnetic. This change from electric to chiral to magnetic with increasing expansion rate, signals that Hubble damping affects charge and current differently, possibly impacting charge more.

\begin{table}
  \begin{center}33
\caption{Scaling exponents of the chirality $|K|$ and total charge and current $Y$. Fitting range for these quantities is $\eta \in [185,435]$. One sigma statistical uncertainties, from the average of ten $2048^3$ simulations, are also reported.}
\label{table2}
  \begin{tabular}{ |c|c|c| }  \hline
Epoch & $\alpha$  & $\beta$  \\    \hline
  $\lambda=1/2$ & $-2.076\pm0.015$ & $-2.307\pm0.004$ \\ 
  $\lambda=2/3$ & N/A & $-2.315\pm0.007$ \\
  $\lambda=3/4$ & $-2.069\pm0.022$ & $-2.230\pm0.009$ \\ \hline
  \end{tabular}
\end{center}
\end{table}

In order to confirm that the networks are mildly electric (radiation), chiral (matter), and mildly magnetic (fast expansion), we can numerically measure the equation of state parameters, either for all degrees of freedom or only for the condensate; the latter is plotted in the bottom part of Fig. \ref{figure2}. For the condensate in the matter era, after the initial timesteps we do find a value consistent with $\omega_\sigma=1/3$, which is exactly what is expected for a freely propagating massless degree of freedom, showing that the condensate degree of freedom is light-like (i.e. chiral). On the other hand, for the radiation era we measure $\omega_\sigma>1/3$, confirming a deviation from chirality. For fast expansion, we can also confirm the value is $\omega_\sigma<1/3$, which is consistent with magnetic strings. Neither the fast expansion or radiation epoch cases show $\omega_\sigma$ approaching chirality at the end of the dynamic range.

\begin{table}
  \begin{center}
\caption{Scaling exponents of the inverse squared coherence lengthscales $\sqrt{1/l_{Q,J}}$. Fitting range for these quantities is $\eta \in [185,512]$. One sigma statistical uncertainties, from the average of ten $2048^3$ simulations, are also reported.}
\label{table3}
  \begin{tabular}{ |c|c|c| }  \hline
Epoch & $\gamma_J$  & $\gamma_Q$  \\    \hline
  $\lambda=1/2$ & $1.010\pm0.053$ & $0.993\pm0.048$ \\ 
  $\lambda=2/3$ & $1.046\pm0.150$ & $1.046\pm0.113$ \\
  $\lambda=3/4$ & $1.051\pm0.157$ & $1.014\pm0.110$ \\ \hline
  \end{tabular}
\end{center}
\end{table}

Lastly we turn our attention to coherence lengthscales $l_{Q,J}$. We posit that if $Y$ and $K$ obey power law decay, then the inverse square root of this lengthscale should exhibit linear growth,
\begin{equation}
\frac{1}{\sqrt{l_{Q,J}}} \propto (\eta - \eta_0)^\gamma ,
\end{equation} 
where $\gamma$ is consistent with unity. Table \ref{table3} confirms that both of these, charge and current, are consistent with linear scaling: the coherence length of charge and current grows as the universe expands. One other detail obvious from the lower left panel of Fig. \ref{figure2} is the relative size of each charge and current lengthscale: at chirality (matter epoch), $l_{Q}^{-1/2} \approx l_{J}^{-1/2}$, in the magnetic regime (fast expansion epoch) the current will vary in lengthscales shorter than the charge and as expected ($l_{Q}^{-1/2} \lessapprox l_{J}^{-1/2}$), whereas in radiation the opposite happens ($l_{Q}^{-1/2} \gtrapprox l_{J}^{-1/2}$). Here we also see how the Universe's expansion affects charge and current differently: faster expansion has a more pronounced effect on charge than it does on current.

\section{\label{conc}Conclusions}

We have extended our multi-GPU simulation of $U(1)\times U(1)$ strings \cite{Correia:2022spe} to explore for the first time the effects of charge and current in the evolution of current-carrying strings, also introducing new numerical diagnostics for the charge and current properties. We have explored the regime of small charges and currents, finding different behaviours in the matter and radiation eras: in the former the networks are chiral, with large-scale scaling and decaying charge and current, while in the radiation era the networks are not chiral.

We emphasize that the model studied in both works is the same, but the two numerically explore different regions of its parameter space. In the former work we had found multitension networks, with clear evidence of scaling for the lightest strings, and also milder evidence for scaling of the bound state segments (which carry a few percent of the network's energy density). In the current one, we found that although the overall network correlation length is clearly approaching scaling, it is less clear that the network velocity is doing the same, at least in the dynamic range we can currently explore with $2048^3$ simulations.

A more robust test of scaling will require larger simulation boxes (or significantly larger numbers of boxes of $2048^3$ size), and is left for future work. The bottleneck in such an analysis is that for it to be fully consistent it should include the behaviour of $K$ and $Y$ (or alternatively $\xi_Q$ and $\xi_J$). As can be seen in the middle panels of fig. \ref{figure2}, these quantities are much harder to measure from simulations than the canonical $\xi$ and $\langle v^2 \rangle$, so a large amount of data is necessary. Moreover, whether or not this small charge and current regime is an unbiased representative of the full parameter space of initial conditions remains to be explored.

A quantitative analytic model for the evolution of generic current and charge carrying string networks, the CVOS model, has been recently developed in \cite{Currents1,Currents2}, who have also studied some of its scaling solutions. Solutions for the particular case of chiral networks were considered in \cite{Oliveira,Auclair}, and an exhaustive classification of all possible scaling solutions, and its dependence on model parameters, has been recently provided in \cite{Thesis}. One salient point of these analyses is that the CVOS model relies on a number of free parameters (describing e.g. loop production, charge and current bias and leakage) which must be measured from simulations, so a detailed comparison of the two approaches is left for future work. Our analysis herein shows that with sufficient computing resources it should be possible to numerically measure these parameters.

Meanwhile, two broad model predictions are that the impact of charges and currents on the network dynamics should be larger in the radiation era than in the matter era and that chiral networks are more likely in the latter. Qualitatively, our simulation results are consistent with both of these, but they also show that an important question in the literature---the extent to which charges and currents prevent or modify the canonical scaling solutions---requires further study. In any case, our results also show that the additional degrees of freedom in the string worldsheet can lead to deviations from the featureless Abelian-Higgs strings behaviour, which are at least temporary but may in some cases be permanent (depending on loss mechanisms), which can induce significant changes in observational signals.

\section*{Acknowledgements}
We acknowledge discussions with Mark Hindmarsh, Ana Achucarro, Paul Shellard and Patrick Peter.

This work was financed by Portuguese funds through FCT (Funda\c c\~ao para a Ci\^encia e a Tecnologia) in the framework of the project 2022.04048.PTDC (Phi in the Sky, DOI 10.54499/2022.04048.PTDC). CJM also acknowledges FCT and POCH/FSE (EC) support through Investigador FCT Contract 2021.01214.CEECIND/CP1658/CT0001 (DOI 10.54499/2021.01214.CEECIND/CP1658/CT0001). 

We acknowledge EuroHPC Joint Undertaking for awarding us access to Karolina at IT4Innovations, Czech Republic. This work was supported by the Ministry of Education, Youth and Sports of the Czech Republic through the e-INFRA CZ (ID:90254).

\bibliographystyle{model1-num-names}
\bibliography{artigo}

\begin{thebibliography}{60}
\expandafter\ifx\csname natexlab\endcsname\relax\def\natexlab#1{#1}\fi
\providecommand{\bibinfo}[2]{#2}
\ifx\xfnm\relax \def\xfnm[#1]{\unskip,\space#1}\fi
\bibitem[{Kibble(1976)}]{Kibble:1976sj}
\bibinfo{author}{T.~W.~B. Kibble},
\newblock \bibinfo{title}{{Topology of Cosmic Domains and Strings}},
\newblock \bibinfo{journal}{J. Phys.} \bibinfo{volume}{A9} (\bibinfo{year}{1976}) \bibinfo{pages}{1387--1398}.
\bibitem[{Jeannerot(1996)}]{Jeannerot1}
\bibinfo{author}{R.~Jeannerot},
\newblock \bibinfo{title}{{A Supersymmetric SO(10) model with inflation and cosmic strings}},
\newblock \bibinfo{journal}{Phys. Rev. D} \bibinfo{volume}{53} (\bibinfo{year}{1996}) \bibinfo{pages}{5426--5436}.
\bibitem[{Jeannerot et~al.(2003)Jeannerot, Rocher, and Sakellariadou}]{Jeannerot2}
\bibinfo{author}{R.~Jeannerot}, \bibinfo{author}{J.~Rocher}, \bibinfo{author}{M.~Sakellariadou},
\newblock \bibinfo{title}{{How generic is cosmic string formation in SUSY GUTs}},
\newblock \bibinfo{journal}{Phys. Rev.} \bibinfo{volume}{D68} (\bibinfo{year}{2003}) \bibinfo{pages}{103514}.
\bibitem[{Ade et~al.(2014)}]{Planck:2013mgr}
\bibinfo{author}{P.~A.~R. Ade}, et~al.,
\newblock \bibinfo{title}{{Planck 2013 results. XXV. Searches for cosmic strings and other topological defects}},
\newblock \bibinfo{journal}{Astron. Astrophys.} \bibinfo{volume}{571} (\bibinfo{year}{2014}) \bibinfo{pages}{A25}.
\bibitem[{Abbott et~al.(2021)}]{LIGO}
\bibinfo{author}{R.~Abbott}, et~al.,
\newblock \bibinfo{title}{{Constraints on Cosmic Strings Using Data from the Third Advanced LIGO\textendash{}Virgo Observing Run}},
\newblock \bibinfo{journal}{Phys. Rev. Lett.} \bibinfo{volume}{126} (\bibinfo{year}{2021}) \bibinfo{pages}{241102}.
\bibitem[{Afzal et~al.(2023)}]{NANOGrav:2023hvm}
\bibinfo{author}{A.~Afzal}, et~al.,
\newblock \bibinfo{title}{{The NANOGrav 15 yr Data Set: Search for Signals from New Physics}},
\newblock \bibinfo{journal}{Astrophys. J. Lett.} \bibinfo{volume}{951} (\bibinfo{year}{2023}) \bibinfo{pages}{L11}.
\bibitem[{Antoniadis et~al.(2023)}]{EPTA}
\bibinfo{author}{J.~Antoniadis}, et~al.,
\newblock \bibinfo{title}{{The second data release from the European Pulsar Timing Array: V. Implications for massive black holes, dark matter and the early Universe}}  (\bibinfo{year}{2023}).
\bibitem[{Colpi et~al.(2024)}]{LISA}
\bibinfo{author}{M.~Colpi}, et~al.,
\newblock \bibinfo{title}{{LISA Definition Study Report}}  (\bibinfo{year}{2024}).
\bibitem[{Weltman et~al.(2020)}]{Weltman:2018zrl}
\bibinfo{author}{A.~Weltman}, et~al.,
\newblock \bibinfo{title}{{Fundamental physics with the Square Kilometre Array}},
\newblock \bibinfo{journal}{Publ. Astron. Soc. Austral.} \bibinfo{volume}{37} (\bibinfo{year}{2020}) \bibinfo{pages}{e002}.
\bibitem[{Kibble(1985)}]{Kibble85}
\bibinfo{author}{T.~W.~B. Kibble},
\newblock \bibinfo{title}{{Evolution of a system of cosmic strings}},
\newblock \bibinfo{journal}{Nucl. Phys. B} \bibinfo{volume}{252} (\bibinfo{year}{1985}) \bibinfo{pages}{227}. \bibinfo{note}{[Erratum: Nucl.Phys.B 261, 750 (1985)]}.
\bibitem[{Bennett(1986)}]{Bennett86}
\bibinfo{author}{D.~P. Bennett},
\newblock \bibinfo{title}{{The evolution of cosmic strings}},
\newblock \bibinfo{journal}{Phys. Rev. D} \bibinfo{volume}{33} (\bibinfo{year}{1986}) \bibinfo{pages}{872}. \bibinfo{note}{[Erratum: Phys.Rev.D 34, 3932 (1986)]}.
\bibitem[{Vachaspati and Vilenkin(1987)}]{Vachaspati87}
\bibinfo{author}{T.~Vachaspati}, \bibinfo{author}{A.~Vilenkin},
\newblock \bibinfo{title}{{Evolution of cosmic networks}},
\newblock \bibinfo{journal}{Phys. Rev. D} \bibinfo{volume}{35} (\bibinfo{year}{1987}) \bibinfo{pages}{1131}.
\bibitem[{Kibble and Copeland(1991)}]{Kibble91}
\bibinfo{author}{T.~W.~B. Kibble}, \bibinfo{author}{E.~J. Copeland},
\newblock \bibinfo{title}{{Evolution of small scale structure on cosmic strings}},
\newblock \bibinfo{journal}{Phys. Scripta T} \bibinfo{volume}{36} (\bibinfo{year}{1991}) \bibinfo{pages}{153--166}.
\bibitem[{Austin et~al.(1993)Austin, Copeland, and Kibble}]{Austin93}
\bibinfo{author}{D.~Austin}, \bibinfo{author}{E.~J. Copeland}, \bibinfo{author}{T.~W.~B. Kibble},
\newblock \bibinfo{title}{{Evolution of cosmic string configurations}},
\newblock \bibinfo{journal}{Phys. Rev. D} \bibinfo{volume}{48} (\bibinfo{year}{1993}) \bibinfo{pages}{5594--5627}.
\bibitem[{Martins and Shellard(1996)}]{Martins:1996jp}
\bibinfo{author}{C.~J. A.~P. Martins}, \bibinfo{author}{E.~P.~S. Shellard},
\newblock \bibinfo{title}{{Quantitative string evolution}},
\newblock \bibinfo{journal}{Phys. Rev.} \bibinfo{volume}{D54} (\bibinfo{year}{1996}) \bibinfo{pages}{2535--2556}.
\bibitem[{Martins and Shellard(2002)}]{Martins:2000cs}
\bibinfo{author}{C.~J. A.~P. Martins}, \bibinfo{author}{E.~P.~S. Shellard},
\newblock \bibinfo{title}{{Extending the velocity dependent one scale string evolution model}},
\newblock \bibinfo{journal}{Phys. Rev. D} \bibinfo{volume}{65} (\bibinfo{year}{2002}) \bibinfo{pages}{043514}.
\bibitem[{Martins(2016)}]{Book}
\bibinfo{author}{C.~J. A.~P. Martins}, \bibinfo{title}{Defect Evolution in Cosmology and Condensed Matter: Quantitative Analysis with the Velocity-Dependent One-Scale Model}, \bibinfo{publisher}{Springer}, \bibinfo{year}{2016}.
\bibitem[{Albrecht and Turok(1985)}]{Albrecht}
\bibinfo{author}{A.~Albrecht}, \bibinfo{author}{N.~Turok},
\newblock \bibinfo{title}{{Evolution of Cosmic Strings}},
\newblock \bibinfo{journal}{Phys. Rev. Lett.} \bibinfo{volume}{54} (\bibinfo{year}{1985}) \bibinfo{pages}{1868--1871}.
\bibitem[{Allen and Shellard(1990)}]{Allen}
\bibinfo{author}{B.~Allen}, \bibinfo{author}{E.~P.~S. Shellard},
\newblock \bibinfo{title}{{Cosmic string evolution: a numerical simulation}},
\newblock \bibinfo{journal}{Phys. Rev. Lett.} \bibinfo{volume}{64} (\bibinfo{year}{1990}) \bibinfo{pages}{119--122}.
\bibitem[{Bennett and Bouchet(1988)}]{Bennett}
\bibinfo{author}{D.~P. Bennett}, \bibinfo{author}{F.~R. Bouchet},
\newblock \bibinfo{title}{{Evidence for a Scaling Solution in Cosmic String Evolution}},
\newblock \bibinfo{journal}{Phys. Rev. Lett.} \bibinfo{volume}{60} (\bibinfo{year}{1988}) \bibinfo{pages}{257}.
\bibitem[{Ringeval et~al.(2007)Ringeval, Sakellariadou, and Bouchet}]{Ringeval}
\bibinfo{author}{C.~Ringeval}, \bibinfo{author}{M.~Sakellariadou}, \bibinfo{author}{F.~Bouchet},
\newblock \bibinfo{title}{{Cosmological evolution of cosmic string loops}},
\newblock \bibinfo{journal}{JCAP} \bibinfo{volume}{02} (\bibinfo{year}{2007}) \bibinfo{pages}{023}.
\bibitem[{Blanco-Pillado et~al.(2011)Blanco-Pillado, Olum, and Shlaer}]{Blanco-Pillado:2011egf}
\bibinfo{author}{J.~J. Blanco-Pillado}, \bibinfo{author}{K.~D. Olum}, \bibinfo{author}{B.~Shlaer},
\newblock \bibinfo{title}{{Large parallel cosmic string simulations: New results on loop production}},
\newblock \bibinfo{journal}{Phys. Rev. D} \bibinfo{volume}{83} (\bibinfo{year}{2011}) \bibinfo{pages}{083514}.
\bibitem[{Vincent et~al.(1998)Vincent, Antunes, and Hindmarsh}]{Vincent}
\bibinfo{author}{G.~Vincent}, \bibinfo{author}{N.~D. Antunes}, \bibinfo{author}{M.~Hindmarsh},
\newblock \bibinfo{title}{{Numerical simulations of string networks in the Abelian Higgs model}},
\newblock \bibinfo{journal}{Phys. Rev. Lett.} \bibinfo{volume}{80} (\bibinfo{year}{1998}) \bibinfo{pages}{2277--2280}.
\bibitem[{Moore et~al.(2002)Moore, Shellard, and Martins}]{Moore}
\bibinfo{author}{J.~N. Moore}, \bibinfo{author}{E.~P.~S. Shellard}, \bibinfo{author}{C.~J. A.~P. Martins},
\newblock \bibinfo{title}{{On the evolution of Abelian-Higgs string networks}},
\newblock \bibinfo{journal}{Phys. Rev. D} \bibinfo{volume}{65} (\bibinfo{year}{2002}) \bibinfo{pages}{023503}.
\bibitem[{Bevis et~al.(2007)Bevis, Hindmarsh, Kunz, and Urrestilla}]{Bevis}
\bibinfo{author}{N.~Bevis}, \bibinfo{author}{M.~Hindmarsh}, \bibinfo{author}{M.~Kunz}, \bibinfo{author}{J.~Urrestilla},
\newblock \bibinfo{title}{{CMB power spectrum contribution from cosmic strings using field-evolution simulations of the Abelian Higgs model}},
\newblock \bibinfo{journal}{Phys. Rev. D} \bibinfo{volume}{75} (\bibinfo{year}{2007}) \bibinfo{pages}{065015}.
\bibitem[{Daverio et~al.(2016)Daverio, Hindmarsh, Kunz, Lizarraga, and Urrestilla}]{Daverio}
\bibinfo{author}{D.~Daverio}, \bibinfo{author}{M.~Hindmarsh}, \bibinfo{author}{M.~Kunz}, \bibinfo{author}{J.~Lizarraga}, \bibinfo{author}{J.~Urrestilla},
\newblock \bibinfo{title}{{Energy-momentum correlations for Abelian Higgs cosmic strings}},
\newblock \bibinfo{journal}{Phys. Rev. D} \bibinfo{volume}{93} (\bibinfo{year}{2016}) \bibinfo{pages}{085014}. \bibinfo{note}{[Erratum: Phys.Rev.D 95, 049903 (2017)]}.
\bibitem[{Hindmarsh et~al.(2017)Hindmarsh, Lizarraga, Urrestilla, Daverio, and Kunz}]{Hindmarsh}
\bibinfo{author}{M.~Hindmarsh}, \bibinfo{author}{J.~Lizarraga}, \bibinfo{author}{J.~Urrestilla}, \bibinfo{author}{D.~Daverio}, \bibinfo{author}{M.~Kunz},
\newblock \bibinfo{title}{{Scaling from gauge and scalar radiation in Abelian Higgs string networks}},
\newblock \bibinfo{journal}{Phys. Rev. D} \bibinfo{volume}{96} (\bibinfo{year}{2017}) \bibinfo{pages}{023525}.
\bibitem[{Correia and Martins(2019)}]{Correia:2019bdl}
\bibinfo{author}{J.~R. C. C.~C. Correia}, \bibinfo{author}{C.~J. A.~P. Martins},
\newblock \bibinfo{title}{{Extending and Calibrating the Velocity dependent One-Scale model for Cosmic Strings with One Thousand Field Theory Simulations}},
\newblock \bibinfo{journal}{Phys. Rev.} \bibinfo{volume}{D100} (\bibinfo{year}{2019}) \bibinfo{pages}{103517}.
\bibitem[{Witten(1985)}]{Witten}
\bibinfo{author}{E.~Witten},
\newblock \bibinfo{title}{{Cosmic Superstrings}},
\newblock \bibinfo{journal}{Phys. Lett. B} \bibinfo{volume}{153} (\bibinfo{year}{1985}) \bibinfo{pages}{243--246}.
\bibitem[{Rybak et~al.(2017)Rybak, Avgoustidis, and Martins}]{Rybak:2017yfu}
\bibinfo{author}{I.~Y. Rybak}, \bibinfo{author}{A.~Avgoustidis}, \bibinfo{author}{C.~J. A.~P. Martins},
\newblock \bibinfo{title}{{Semianalytic calculation of cosmic microwave background anisotropies from wiggly and superconducting cosmic strings}},
\newblock \bibinfo{journal}{Phys. Rev. D} \bibinfo{volume}{96} (\bibinfo{year}{2017}) \bibinfo{pages}{103535}. \bibinfo{note}{[Erratum: Phys.Rev.D 100, 049901 (2019)]}.
\bibitem[{Lilley et~al.(2010)Lilley, Di~Marco, Martin, and Peter}]{Peter3}
\bibinfo{author}{M.~Lilley}, \bibinfo{author}{F.~Di~Marco}, \bibinfo{author}{J.~Martin}, \bibinfo{author}{P.~Peter},
\newblock \bibinfo{title}{{Nonabelian Bosonic Currents in Cosmic Strings}},
\newblock \bibinfo{journal}{Phys. Rev. D} \bibinfo{volume}{82} (\bibinfo{year}{2010}) \bibinfo{pages}{023510}.
\bibitem[{Gangui et~al.(1999)Gangui, Peter, and Gunzig}]{Peter1}
\bibinfo{author}{A.~Gangui}, \bibinfo{author}{P.~Peter}, \bibinfo{author}{E.~Gunzig},
\newblock \bibinfo{title}{{Fermionic currents in cosmic strings}},
\newblock \bibinfo{journal}{Int. J. Theor. Phys.} \bibinfo{volume}{38} (\bibinfo{year}{1999}) \bibinfo{pages}{205--216}.
\bibitem[{Davis and Shellard(1989)}]{Davis:1988ij}
\bibinfo{author}{R.~L. Davis}, \bibinfo{author}{E.~P.~S. Shellard},
\newblock \bibinfo{title}{{COSMIC VORTONS}},
\newblock \bibinfo{journal}{Nucl. Phys. B} \bibinfo{volume}{323} (\bibinfo{year}{1989}) \bibinfo{pages}{209--224}.
\bibitem[{Peter(1993)}]{Peter:1993pz}
\bibinfo{author}{P.~Peter},
\newblock \bibinfo{title}{{Electromagnetically supported cosmic string loops}},
\newblock \bibinfo{journal}{Phys. Lett. B} \bibinfo{volume}{298} (\bibinfo{year}{1993}) \bibinfo{pages}{60--62}.
\bibitem[{Auclair et~al.(2021)Auclair, Peter, Ringeval, and Steer}]{Auclair:2020wse}
\bibinfo{author}{P.~Auclair}, \bibinfo{author}{P.~Peter}, \bibinfo{author}{C.~Ringeval}, \bibinfo{author}{D.~Steer},
\newblock \bibinfo{title}{{Irreducible cosmic production of relic vortons}},
\newblock \bibinfo{journal}{JCAP} \bibinfo{volume}{03} (\bibinfo{year}{2021}) \bibinfo{pages}{098}.
\bibitem[{Vilenkin and Vachaspati(1987)}]{Vilenkin:1986zz}
\bibinfo{author}{A.~Vilenkin}, \bibinfo{author}{T.~Vachaspati},
\newblock \bibinfo{title}{{Electromagnetic Radiation from Superconducting Cosmic Strings}},
\newblock \bibinfo{journal}{Phys. Rev. Lett.} \bibinfo{volume}{58} (\bibinfo{year}{1987}) \bibinfo{pages}{1041--1044}.
\bibitem[{Vachaspati(2010)}]{Vachaspati:2009kq}
\bibinfo{author}{T.~Vachaspati},
\newblock \bibinfo{title}{{Cosmic Rays from Cosmic Strings with Condensates}},
\newblock \bibinfo{journal}{Phys. Rev. D} \bibinfo{volume}{81} (\bibinfo{year}{2010}) \bibinfo{pages}{043531}.
\bibitem[{Th\'eriault et~al.(2021)Th\'eriault, Mirocha, and Brandenberger}]{Theriault:2021mrq}
\bibinfo{author}{R.~Th\'eriault}, \bibinfo{author}{J.~T. Mirocha}, \bibinfo{author}{R.~Brandenberger},
\newblock \bibinfo{title}{{Global 21cm absorption signal from superconducting cosmic strings}},
\newblock \bibinfo{journal}{JCAP} \bibinfo{volume}{10} (\bibinfo{year}{2021}) \bibinfo{pages}{046}.
\bibitem[{Martins et~al.(2021{\natexlab{a}})Martins, Peter, Rybak, and Shellard}]{Currents1}
\bibinfo{author}{C.~J. A.~P. Martins}, \bibinfo{author}{P.~Peter}, \bibinfo{author}{I.~Y. Rybak}, \bibinfo{author}{E.~P.~S. Shellard},
\newblock \bibinfo{title}{{Generalized velocity-dependent one-scale model for current-carrying strings}},
\newblock \bibinfo{journal}{Phys. Rev. D} \bibinfo{volume}{103} (\bibinfo{year}{2021}{\natexlab{a}}) \bibinfo{pages}{043538}.
\bibitem[{Martins et~al.(2021{\natexlab{b}})Martins, Peter, Rybak, and Shellard}]{Currents2}
\bibinfo{author}{C.~J. A.~P. Martins}, \bibinfo{author}{P.~Peter}, \bibinfo{author}{I.~Y. Rybak}, \bibinfo{author}{E.~P.~S. Shellard},
\newblock \bibinfo{title}{{Charge-velocity-dependent one-scale linear model}},
\newblock \bibinfo{journal}{Phys. Rev. D} \bibinfo{volume}{104} (\bibinfo{year}{2021}{\natexlab{b}}) \bibinfo{pages}{103506}.
\bibitem[{Peter(1992{\natexlab{a}})}]{Peter92a}
\bibinfo{author}{P.~Peter},
\newblock \bibinfo{title}{{Superconducting cosmic string: Equation of state for space - like and time - like current in the neutral limit}},
\newblock \bibinfo{journal}{Phys. Rev. D} \bibinfo{volume}{45} (\bibinfo{year}{1992}{\natexlab{a}}) \bibinfo{pages}{1091--1102}.
\bibitem[{Peter(1992{\natexlab{b}})}]{Peter92b}
\bibinfo{author}{P.~Peter},
\newblock \bibinfo{title}{{Influence of the electric coupling strength in current carrying cosmic strings}},
\newblock \bibinfo{journal}{Phys. Rev. D} \bibinfo{volume}{46} (\bibinfo{year}{1992}{\natexlab{b}}) \bibinfo{pages}{3335--3349}.
\bibitem[{Carter and Peter(1995)}]{Carter95}
\bibinfo{author}{B.~Carter}, \bibinfo{author}{P.~Peter},
\newblock \bibinfo{title}{{Supersonic string models for Witten vortices}},
\newblock \bibinfo{journal}{Phys. Rev. D} \bibinfo{volume}{52} (\bibinfo{year}{1995}) \bibinfo{pages}{1744--1748}.
\bibitem[{Carter(2001)}]{Carter01}
\bibinfo{author}{B.~Carter},
\newblock \bibinfo{title}{{Essentials of classical brane dynamics}},
\newblock \bibinfo{journal}{Int. J. Theor. Phys.} \bibinfo{volume}{40} (\bibinfo{year}{2001}) \bibinfo{pages}{2099--2130}.
\bibitem[{Copeland and Saffin(2005)}]{Copeland05}
\bibinfo{author}{E.~J. Copeland}, \bibinfo{author}{P.~M. Saffin},
\newblock \bibinfo{title}{{On the evolution of cosmic-superstring networks}},
\newblock \bibinfo{journal}{JHEP} \bibinfo{volume}{11} (\bibinfo{year}{2005}) \bibinfo{pages}{023}.
\bibitem[{Lemperiere and Shellard(2003)}]{Lemperiere:2003yt}
\bibinfo{author}{Y.~Lemperiere}, \bibinfo{author}{E.~P.~S. Shellard},
\newblock \bibinfo{title}{{Vorton existence and stability}},
\newblock \bibinfo{journal}{Phys. Rev. Lett.} \bibinfo{volume}{91} (\bibinfo{year}{2003}) \bibinfo{pages}{141601}.
\bibitem[{Battye et~al.(2022)Battye, Cotterill, and Pearson}]{Battye:2021kbd}
\bibinfo{author}{R.~A. Battye}, \bibinfo{author}{S.~J. Cotterill}, \bibinfo{author}{J.~A. Pearson},
\newblock \bibinfo{title}{{A detailed study of the stability of vortons}},
\newblock \bibinfo{journal}{JHEP} \bibinfo{volume}{04} (\bibinfo{year}{2022}) \bibinfo{pages}{005}.
\bibitem[{Battye and Cotterill(2023)}]{Battye:2022mxi}
\bibinfo{author}{R.~A. Battye}, \bibinfo{author}{S.~J. Cotterill},
\newblock \bibinfo{title}{{Pinching instabilities in superconducting cosmic strings}},
\newblock \bibinfo{journal}{Phys. Rev. D} \bibinfo{volume}{107} (\bibinfo{year}{2023}) \bibinfo{pages}{063534}.
\bibitem[{Correia and Martins(2021)}]{Correia:2020yqg}
\bibinfo{author}{J.~R. C. C.~C. Correia}, \bibinfo{author}{C.~J. A.~P. Martins},
\newblock \bibinfo{title}{{Abelian\textendash{}Higgs cosmic string evolution with multiple GPUs}},
\newblock \bibinfo{journal}{Astron. Comput.} \bibinfo{volume}{34} (\bibinfo{year}{2021}) \bibinfo{pages}{100438}.
\bibitem[{Correia and Martins(2022)}]{Correia:2022spe}
\bibinfo{author}{J.~R. C. C.~C. Correia}, \bibinfo{author}{C.~J. A.~P. Martins},
\newblock \bibinfo{title}{{Multitension strings in high-resolution U(1)\texttimes{}U(1) simulations}},
\newblock \bibinfo{journal}{Phys. Rev. D} \bibinfo{volume}{106} (\bibinfo{year}{2022}) \bibinfo{pages}{043521}.
\bibitem[{Saffin(2005)}]{Saffin:2005cs}
\bibinfo{author}{P.~M. Saffin},
\newblock \bibinfo{title}{{A Practical model for cosmic (p,q) superstrings}},
\newblock \bibinfo{journal}{JHEP} \bibinfo{volume}{09} (\bibinfo{year}{2005}) \bibinfo{pages}{011}.
\bibitem[{Wilson(1974)}]{PhysRevD.10.2445}
\bibinfo{author}{K.~G. Wilson},
\newblock \bibinfo{title}{Confinement of quarks},
\newblock \bibinfo{journal}{Phys. Rev. D} \bibinfo{volume}{10} (\bibinfo{year}{1974}) \bibinfo{pages}{2445--2459}.
\bibitem[{Press et~al.(1989)Press, Ryden, and Spergel}]{PRS}
\bibinfo{author}{W.~H. Press}, \bibinfo{author}{B.~S. Ryden}, \bibinfo{author}{D.~N. Spergel},
\newblock \bibinfo{title}{{Dynamical Evolution of Domain Walls in an Expanding Universe}},
\newblock \bibinfo{journal}{Astrophys. J.} \bibinfo{volume}{347} (\bibinfo{year}{1989}) \bibinfo{pages}{590--604}.
\bibitem[{Lizarraga and Urrestilla(2016)}]{Lizarraga:2016hpd}
\bibinfo{author}{J.~Lizarraga}, \bibinfo{author}{J.~Urrestilla},
\newblock \bibinfo{title}{{Survival of pq-superstrings in field theory simulations}},
\newblock \bibinfo{journal}{JCAP} \bibinfo{volume}{04} (\bibinfo{year}{2016}) \bibinfo{pages}{053}.
\bibitem[{Fujikura et~al.(2023)Fujikura, Li, and Yamaguchi}]{Fujikura:2023lil}
\bibinfo{author}{K.~Fujikura}, \bibinfo{author}{S.~Li}, \bibinfo{author}{M.~Yamaguchi},
\newblock \bibinfo{title}{{Interactions between several types of cosmic strings}},
\newblock \bibinfo{journal}{JHEP} \bibinfo{volume}{12} (\bibinfo{year}{2023}) \bibinfo{pages}{115}.
\bibitem[{zen(2024)}]{zenodo}
\bibinfo{title}{{Superconducting string networks animation}}, \bibinfo{howpublished}{\url{https://doi.org/10.5281/zenodo.10845839}}, \bibinfo{year}{2024}.
\bibitem[{Kajantie et~al.(1998)Kajantie, Karjalainen, Laine, Peisa, and Rajantie}]{Kajantie:1998bg}
\bibinfo{author}{K.~Kajantie}, \bibinfo{author}{M.~Karjalainen}, \bibinfo{author}{M.~Laine}, \bibinfo{author}{J.~Peisa}, \bibinfo{author}{A.~Rajantie},
\newblock \bibinfo{title}{{Thermodynamics of gauge invariant U(1) vortices from lattice Monte Carlo simulations}},
\newblock \bibinfo{journal}{Phys. Lett. B} \bibinfo{volume}{428} (\bibinfo{year}{1998}) \bibinfo{pages}{334--341}.
\bibitem[{Oliveira et~al.(2012)Oliveira, Avgoustidis, and Martins}]{Oliveira}
\bibinfo{author}{M.~F. Oliveira}, \bibinfo{author}{A.~Avgoustidis}, \bibinfo{author}{C.~J. A.~P. Martins},
\newblock \bibinfo{title}{{Cosmic string evolution with a conserved charge}},
\newblock \bibinfo{journal}{Phys. Rev. D} \bibinfo{volume}{85} (\bibinfo{year}{2012}) \bibinfo{pages}{083515}.
\bibitem[{Auclair et~al.(2023)Auclair, Blasi, Brdar, and Schmitz}]{Auclair}
\bibinfo{author}{P.~Auclair}, \bibinfo{author}{S.~Blasi}, \bibinfo{author}{V.~Brdar}, \bibinfo{author}{K.~Schmitz},
\newblock \bibinfo{title}{{Gravitational waves from current-carrying cosmic strings}},
\newblock \bibinfo{journal}{JCAP} \bibinfo{volume}{04} (\bibinfo{year}{2023}) \bibinfo{pages}{009}.
\bibitem[{Pimenta(2023)}]{Thesis}
\bibinfo{author}{F.~C. N.~Q. Pimenta}, \bibinfo{title}{Analytical solutions for the evolution of current carrying cosmic strings}, Master's thesis, University of Porto, \bibinfo{year}{2023}.

\end{thebibliography}
\end{document}